 \documentclass[manuscript]{emulateapj}

\slugcomment{to appear in the Astrophysical Journal, Letters}

\shorttitle{Delay Time Distribution}
\shortauthors{Hachisu et al.}

\begin{document}

\title{The Delay Time Distribution of Type I\lowercase{a}
Supernovae and the Single Degenerate Model}


\author{Izumi Hachisu}
\affil{Department of Earth Science and Astronomy,
College of Arts and Sciences, University of Tokyo,
Komaba 3-8-1, Meguro-ku, Tokyo 153-8902, Japan}
\email{hachisu@ea.c.u-tokyo.ac.jp}

\author{Mariko Kato}
\affil{Department of Astronomy, Keio University,
Hiyoshi 4-1-1, Kouhoku-ku, Yokohama 223-8521, Japan}
\email{mariko@educ.cc.keio.ac.jp}

\and

\author{Ken'ichi Nomoto}
\affil{Institute for the Physics and Mathematics of the Universe,
University of Tokyo, Kashiwanoha 5-1-5, Kashiwa, Chiba 277-8568,
and Department of Astronomy \& Research Center for the Early
Universe, University of Tokyo, Hongo 7-3-1, Bunkyo-ku, Tokyo 113-0033,
Japan}
\email{nomoto@astron.s.u-tokyo.ac.jp}



\begin{abstract}
We present a theoretical delay time distribution (DTD) of Type Ia
supernovae on the basis of our new evolutionary models of single
degenerate (SD) progenitor systems.  Our model DTD has almost a
featureless power law shape ($\propto t^{-n}$ with $n \approx 1$) for
the delay time from $t \sim$ 0.1 to 10 Gyr.  This is in good agreement
with the recent direct measurement of DTD.  The observed featureless
property of the DTD has been suggested to be favorable for the double
degenerate (DD) scenario but not for the SD scenario.  If the mass
range of the companion star to the white dwarf (WD) were too narrow
in the SD model, its DTD would be too limited around the companion's
main-sequence lifetime to be consistent with the observed DTD.
However, this is not the case in our SD model that consists of
the two channels of WD + RG (red giant) and WD + MS (main-sequence
star).  In these channels, the companion stars have a mass range of
$\sim 0.9-3~M_\sun$ (WD$+$RG) and $\sim 2- 6~M_\sun$ (WD$+$MS).  The
combined mass range is wide enough to yield the featureless DTD.  We
emphasize that the SD scenario should include two important processes:
the optically thick winds from the mass-accreting WD and the
mass-stripping from the companion star by the WD wind.
\end{abstract}




\section{Introduction}
Type Ia supernovae (SNe Ia) play the important roles in astrophysics
as a standard candle to measure cosmological distances as well as
the production site of a large part of iron group elements.
However, the nature of SN Ia progenitors has not been
clarified yet \citep[e.g.,][]{hil00, nom97, nom00, liv00}.
It has been commonly agreed that the exploding star
is a carbon-oxygen white dwarf (C+O WD) and the observed
features of SNe Ia are better explained by the Chandrasekhar mass
model than the sub-Chandrasekhar mass model.
However, there has been no clear observational indication
as to how the WD mass gets close enough to the Chandrasekhar mass for
carbon ignition; i.e., whether the WD accretes H/He-rich matter from
its binary companion [single degenerate (SD) scenario] or two C+O WDs
merge [double degenerate (DD) scenario].

It has been suggested that SNe Ia have a wide range of delay time
from $t < 0.1$ Gyr to $t > 10$ Gyr \citep[e.g.,][]{man06}.
Here the delay time, $t$, is defined as the age at the explosion
of the SN Ia progenitor from its birth.
According to \citet{man06}, the present observational data of SNe Ia
are best matched by a bimodal population of the progenitors, in which
about 50 percent of SNe Ia explode soon after their stellar birth
at the delay time of $t \sim 0.1$ Gyr,
while the remaining 50 percent have a much wider distribution
of the delay time around $t \sim$ 3 Gyr.
If the delay time distribution (DTD) of SNe Ia is observationally
obtained, we are able to preclude some models that are inconsistent
with the DTD.

Recently, the direct measurement of the DTD has been reported
by \citet{tot08}.  Their DTD shows a featureless power law 
($\propto t^{-n}$, $n \approx 1$) between $t \sim 0.1$
and $\sim 10$ Gyr.  On the basis of their results,
they argued that the DTD strongly supports the DD scenario of
SN Ia progenitors mainly because the featureless power law 
distribution is in good agreement with the prediction of
the DD scenario.  They also concluded that the SD scenario
is not well supported mainly because some ``detailed''
binary population synthesis codes predict prominent peaks 
in the DTD at characteristic time scales 
\citep[e.g.,][]{bel05, men08, rui98, yun00},
although some ``simple'' SD models with 
simplified treatments of binary evolution 
have broad DTD shapes similar to the observed DTD
\citep[e.g.,][]{gre05, mat06, kob08}.  

In the SD scenario, the DTD is closely related to the main-sequence
(MS) lifetime of the companion star and thus the initial
mass of the companion (secondary), $M_{2,0}$.
This is because the mass transfer from the companion to the WD
starts when the companion evolves off the MS and expands
to fill its Roche lobe.
The mass-accreting WD quickly grows to the Chandrasekhar mass limit
in $\sim 10^6$ yr \citep[see e.g.,][]{hkn99, hknu99, hkn08},
which is much shorter than the age of the binary system
$t > 10^8$ yr.
The featureless power law ($\propto t^{-n}$) of the DTD between 
$t \sim 0.1$ and $\sim 10$ Gyr
requires a very broad mass distribution of the companion
from $M_{2,0} \sim 0.9$ to $\sim 6 ~M_\sun$.

In the ``detailed'' SD models,
they follow each binary evolution including many binary
evolutionary processes and, as a result, the mass of the companion
that can produce an SN Ia is constrained to a certain range, e.g.,
$M_{2,0} \approx 2-3.5~M_\sun$ in \citet{men08}.
Therefore, the delay time
is also constrained to a narrow range of $t \sim 0.3-1.2$ Gyr,
which is inconsistent with the observed DTD.

In some ``simple'' SD models, a mass range of the companion is
assumed {\it a priori} to be $M_{2,0}= 0.8-8~M_\sun$ 
simply from the condition
of $M_{2,0} < M_{1,0}$ without taking into account the constraints on
the mass accretion rate onto the WD and thus on the separation between
the WD and the companion \citep{gre05}.

In the present paper, we show that the SD model with taking into
account the ``detailed'' binary evolution is viable against the
observed DTD.  Actually, the required broad distribution of the
companion mass, $M_{2,0} \sim 0.9 - 6 ~M_\sun$ has already been
predicted by a recent ``detailed'' progenitor model of SNe Ia
\citep{hkn08}.  However, Hachisu et al. did not present
any DTDs mainly because at that time there were no observational
data to compare with the theoretical results. 
Here we present DTDs on the basis of the new ``detailed'' SD models
\citep{hkn08} in \S\S 2 and 3 and compare with the observation
\citep{tot08} in \S 4.



\begin{figure}
\epsscale{1.15}
\plotone{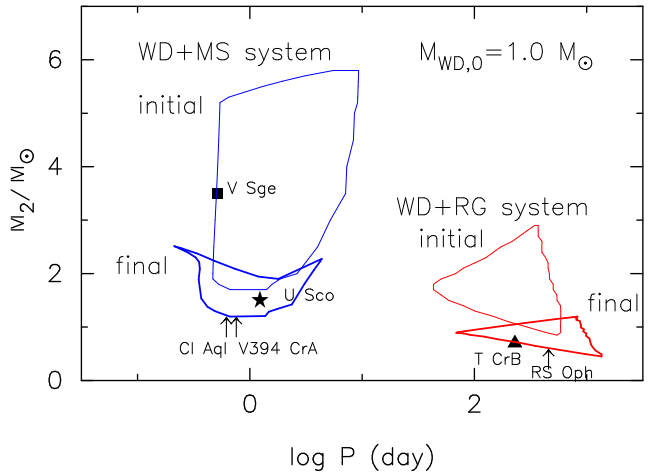}
\caption{
The regions that produce SNe Ia are plotted in the $\log P - M_2$
(orbital period -- secondary mass) plane for the white dwarf and
main-sequence star (WD + MS) system ({\it left}) and the white dwarf
and red giant (WD + RG) system ({\it right}).
Here we assume the metallicity of $Z=0.02$ and 
the initial white dwarf mass of $M_{\rm WD, 0}= 1.0 ~M_\sun$.
The initial system inside the region encircled by a thin solid line
(labeled ``initial'') increases its WD mass up to the critical
mass ($M_{\rm Ia}= 1.38 M_\sun$) for the SN Ia 
explosion, the regions of which are encircled by a thick solid line
(labeled ``final'').  Currently known positions of the recurrent novae
and supersoft X-ray sources are indicated by a star mark ($\star$)
for U Sco \citep[e.g.,][]{hkkm00}, a triangle for T CrB 
\citep[e.g.,][]{bel98}, a square for V Sge \citep{hac03kc},
but by arrows for the other three recurrent novae, V394 CrA,
CI Aql, and RS Oph, because the mass of the companion is not yet
available explicitly.  Two subclasses of the recurrent novae, the
U Sco type and the RS Oph type, correspond to the WD + MS
channel and the WD + RG channel of SNe Ia, respectively.
\label{zregevl10_strip_ms_rg}}
\end{figure}

\section{Mass-Stripping Effect and Binary Evolution}
We start the binary evolution from the zero-age MS.
Unless the initial separation of the binary components is too close,
the more massive (primary with the mass of $M_{1,0}$)
component evolves to a red giant star (with a helium core)
or an AGB star (with a C+O core) and fills its Roche lobe.
Subsequent mass transfer from the primary to the secondary is
rapid enough to form a common envelope.  The binary separation
shrinks greatly owing to the mass and angular momentum losses
from the binary system during the first common envelope evolution.
The hydrogen-rich envelope of the primary component
is stripped away and the primary becomes a helium star
or a C+O WD.  The helium star further evolves to a C+O WD
after a large part of helium is exhausted by core-helium-burning.
Thus we have a binary pair of the C+O WD and the secondary star
that is an MS star; the mass of the secondary star, $M_2$, is
still close to $M_{2,0}$, because the accreted mass
during the common envelope phase is negligibly small.  

After the secondary evolves to fill its Roche lobe,
the WD accretes mass from the secondary and
grows to the critical mass ($M_{\rm Ia}= 1.38~M_\sun$) and
explodes as an SN Ia if the initial binary orbital period ($P_0$)
and the initial mass of the secondary ($M_{2,0}$)
are inside the regions (labeled ``initial'') 
shown in Figure \ref{zregevl10_strip_ms_rg}.
There are two separate regions; one is
for binaries consisting of a white dwarf and a main-sequence
star (WD + MS) and the other is binaries consisting
of a white dwarf and a red giant (WD + RG).
Here the metallicity and the initial white dwarf mass are assumed
to be $Z=0.02$ and $M_{\rm WD,0}= 1.0 ~M_\sun$.
Note that the WD with $M_{\rm WD,0}= 1.0 ~M_\sun$ for $Z=0.02$
forms from the primary star of $M_{1,0} \sim 7 ~M_\sun$ \citep{ume99}.





\begin{figure*}
\plotone{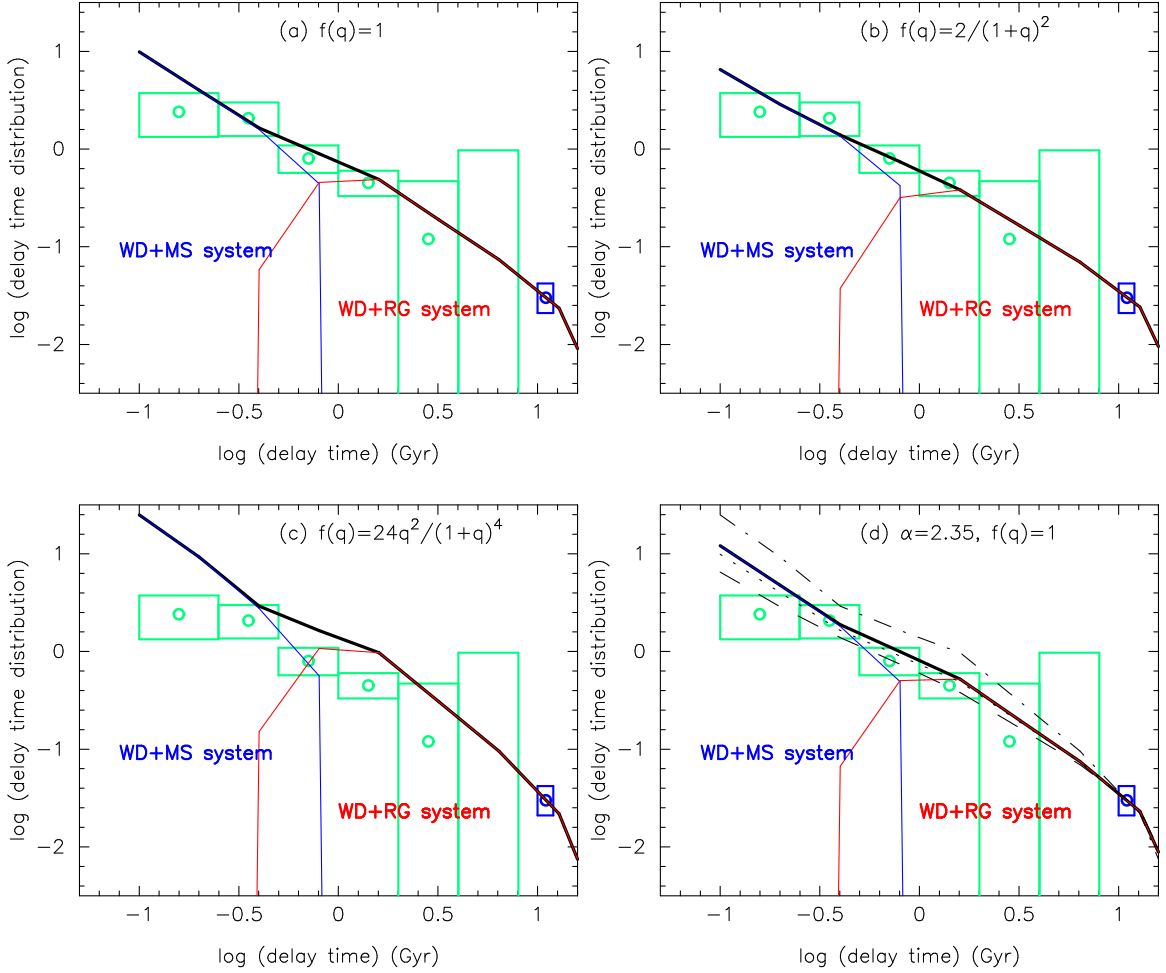}
\caption{
The delay time distributions (DTDs) for our single degenerate
(SD) models of SNe Ia.  The ordinate is the DTD in units
of per century and per $10^{10} L_{K, \sun}$. 
Open circles with an open box are the observational DTD taken from
\citet{tot08} (for $< 10$ Gyr) and \citet{man05} (at 11 Gyr).
Each open box indicates a one-sigma error of each measurement.
Throughout all figures from (a) to (d), thick solid lines indicate
the total DTD coming both from the WD + MS and
WD + RG systems.  Each contribution from each channel is separately
shown by thin solid lines.  The data of our DTD only for (a) are
tabulated in Table \ref{sn1a_delay_time_z02_const}.
The binary mass ratio distribution and the initial mass function
power are assumed to be (a) $f(q)=1$ and $\alpha = 2.5$, 
(b) $f(q)= 2/(1+q)^2$ and $\alpha = 2.5$, where
a factor of 2 is a normalization factor
for $\int_0^1 f(q) d q =1$, and
${\nu}_{\rm WD + MS} = 0.0038$~yr$^{-1}$ 
and ${\nu}_{\rm WD + RG} = 0.0038$~yr$^{-1}$ in our Galaxy,
(c) $f(q)= 24 q^2 /(1+q)^4$ \citep{gre83} and $\alpha = 2.5$, where
a factor of 24 is a normalization factor
for $\int_0^1 f(q) d q =1$, and ${\nu}_{\rm WD + MS} = 0.0034$~yr$^{-1}$
and ${\nu}_{\rm WD + RG} = 0.0025$~yr$^{-1}$ in our Galaxy, and
(d) $f(q)=1$ and $\alpha = 2.35$. 
Dotted line: same as that in Fig. \ref{delay_time_4figures}a,
dashed line: same as that in Fig. \ref{delay_time_4figures}b,
dash-dotted line: same as that in Fig. \ref{delay_time_4figures}c.
\label{delay_time_4figures}}
\end{figure*}

Here we emphasize the effects of two important processes in the
binary evolutions.  The first one is the accretion wind evolution.
Mass-accreting WDs blow strong winds when
the mass transfer rate onto the WD exceeds the critical rate of
${\dot M}_{\rm cr} \sim 1 \times 10^{-6} M_\sun$~yr$^{-1}$
\citep{hkn96}.  The angular momentum taken away by this fast wind
is much smaller than the orbital angular momentum, 
so that the separation of the binary does not shrink \citep{hkn99}.
If we ignore this wind effect, almost 
all the binaries would undergo a common envelope phase and 
merge (the WD + MS systems) or form a double degenerate system
(the WD + RG systems) due to a large amount of angular momentum loss.
None of them becomes an SN Ia through the SD
channel \citep[e.g.,][]{hkn99}.

The second one is the mass-stripping from the secondary surface
by the WD winds.  This attenuates the mass transfer rate 
from the secondary to the WD, so that the binary can
avoid the formation of a common envelope even for a rather massive
secondary \citep{hkn99, hkn08}.  In our results
in Figure \ref{zregevl10_strip_ms_rg}, this mass-stripping 
effect is critically important.

Our results showed that the ``initial'' region of WD + MS 
systems extends up to such a massive ($M_{2,0} \sim 5-6 ~M_\sun$)
secondary, which consists of a very young population of SNe Ia
with such a short delay time as $t \lesssim 0.1$~Gyr.
On the other hand, the WD + RG systems with a less massive RG
($M_{2,0} \sim 0.9-1.0 ~M_\sun$) consist of a very old population of
SNe Ia of $t \gtrsim 10$ Gyr.

\section{Delay Time Distribution}
The birth rate of SNe Ia in our Galaxy is expressed as
\begin{equation}
\nu = 0.2 ~\int \int \int_D {{d M_{1,0}} \over {{(M_{1,0})^{2.5}}}}
f(q)~ d q~ d \log ~a \quad {\rm yr}^{-1},
\label{realization_frequency}
\end{equation}
along equation (1) of \citet{ibe84}, 
where $q= M_{\rm 2,0}/M_{\rm 1,0}$, $a$, and $M_{1,0}$ are
the mass ratio, the separation, and the primary mass
in solar mass units, respectively, at the birth of a binary
and $D$ is the SN Ia region in the $(M_{1,0}, q, a)$-space.
Dividing the three-dimensional space of $(M_{1,0}, q, a)$ into
grids of $(M_i, q_j, a_k)$, 
we follow the binary evolution starting from the initial 
state of $(M_i, q_j, a_k)$ and obtain $\nu$ for
the binaries which finally explode as SNe Ia.  At the same time,
we record the delay time of $t$ for each $(M_i, q_j, a_k)$.
In our calculation, the SN Ia region of $D$ consists
of these SN Ia grid points.
This procedure can easily be done when we know the ``initial''
SN Ia regions for both the WD + MS and WD + RG systems as shown
in Figure \ref{zregevl10_strip_ms_rg}.

In our previous works as seen in Figure 4 of \citet{hkn08} for the WD
+ MS system and Figure 12 of \citet{hkn99} for the WD + RG system,
we have obtained such SN Ia regions for different initial WD masses,
$M_{\rm WD,0}= 0.7$, 0.8, 0.9, 1.0, and $1.1 ~M_\sun$, which formed
from the primary star of masses $M_{1,0} \sim 4$, 5, 6, 7,
and $8 ~M_\sun$ at the birth, respectively,
for $Z = 0.02$ \citep{ume99}.
Using these results, we have estimated the SN Ia birth rate in our
Galaxy as $\nu_{\rm WD + MS} = 0.0035$~yr$^{-1}$ and $\nu_{\rm WD + RG} =
0.0032$~yr$^{-1}$, respectively \citep[see][for more detail]{hkn99,
hknu99, hkn08}.  Here we assume $f(q)=1$ along \citet{ibe84}.
Therefore, the relative ratio of the SNe from the WD + MS progenitors
to those from the WD + RG progenitors is $r_{\rm MS/RG}=1.1$.

Now we estimate the DTD of SNe Ia forming from the WD + MS
and WD + RG systems, by integrating only the initial sets of
$(M_i, q_j, a_k)$ having the delay time between $t - \Delta t$
and $t + \Delta t$, as
\begin{equation}
{\rm DTD}(t) \propto {{1}\over {2 \Delta t}}
\int \int \int_{t-\Delta t}^{t+\Delta t} 
{{d M_{1,0}} \over {{(M_{1,0})}^{2.5}}}
f(q)~ d q~ d \log ~a,
\label{delay_time_distribution}
\end{equation}
where we adopt $f(q)=1$ and 10 bins of the delay time at $t=
0.05$, 0.1, 0.2, 0.4, 0.8, 1.6, 3.2, 6.4, 12.8, and 25.6 Gyr.
The resultant DTD  as summarized
in Table \ref{sn1a_delay_time_z02_const}, 
is not normalized, thus being scale-free.
We normalize our DTD by fitting the value to the observation
at 11 Gyr as shown in Figure \ref{delay_time_4figures}a.  
Our DTD shows a featureless power law ($\propto t^{-n}$,
$n \approx 1$) from 0.1 to 12 Gyr, which is in good agreement
with Totani et al.'s (2008) DTD and the data at 11 Gyr
by \citet{man05}.  Here, we assume
$H_0=70$~km~s$^{-1}$Mpc$^{-1}$ for the Hubble constant.
Thus the DTD on the basis of our ``detailed`` SD model of
SN Ia progenitors is consistent with the observation.
It should be noticed that we plotted the number ratio of SNe Ia
against the delay time in Figure 12 of \citet{hkn08}, which clearly
shows a bimodality of SN Ia progenitors, but the number ratio itself
is not equal to the DTD of SNe Ia because the DTD is time-derivative
of the number ratio.

Here, for instructive purposes,
we derive an approximate power law of our DTD.
The main-sequence lifetime of the secondary, $t_2$,
can be estimated
as $t_2 \propto M_{2,0} / L_{2,0} \propto (M_{2,0})^{1-m}$,
where the mass-luminosity relation
at the zero-age main-sequence is approximately written
as $L_{2,0} \propto (M_{2,0})^m$ with
$m= 3.5$ for $M_{2,0} = 3 - 7 ~M_\sun$.
The appropriate range of initial 
separation ($\Delta \log a$) is roughly proportional to the range
of orbital period, i.e., $\Delta \log a \approx (2/3) \Delta \log P$,
so the area of $\Delta M_{2,0}~ \Delta \log ~a$
is calculated approximately from the SN Ia region
in the $\log P - M_2$ plane for a given
$M_{1,0}$ like in Figure \ref{zregevl10_strip_ms_rg}.
This area becomes narrower
as the initial WD mass, $M_{\rm WD,0}$, decreases
like Figure 4 of \citet{hkn08}.
The initial WD mass is closely related to the initial primary 
mass, $M_{1,0}$, so that
we numerically obtain the approximate relation of 
\begin{equation}
\int \int_D d M_{2,0} ~d \log ~a \propto (M_{1,0})^{2.5},
\label{m10_power}
\end{equation}
for the fixed $M_{1,0}$ between $6 < M_{1,0} < 9$.
We also numerically obtain a similar relation of
\begin{equation}
\int \int_D (M_{1,0})^{-3.5} d M_{1,0} ~d \log ~a \propto
(M_{2,0})^{-1.0},
\label{m20_power}
\end{equation}
for the fixed $M_{2,0}$ between $3 < M_{2,0} < 6$.
Then the DTD can be approximated as
\begin{eqnarray}
{\rm DTD}(t) & \propto & {{1}\over {2 \Delta t}}
\int_{t - \Delta t}^{t + \Delta t} d M_{2,0}
\int \int {{d M_{1,0}} \over {(M_{1,0})^{3.5}}} d \log a \cr
& \propto & {{1}\over {2 \Delta t}} \int_{t - \Delta t}^{t + \Delta t}
 {{d M_{2,0}} \over {M_{2,0}}}
\propto {{1}\over {2 \Delta t}} \int_{t-\Delta t}^{t+\Delta t}
{{d t_2} \over {t_2}} \propto t^{-1},
\label{delay_time_simple_estimate_MS}
\end{eqnarray}
for $\Delta t \ll t_2$.  Here, we assume $f(q)=1$ and
$M_{2,0} \propto (t_2)^{-1/2.5}$.  Strictly speaking, the power
of $M_{2,0}$ in equation (\ref{m20_power}) is somewhere between
$-1.0$ and $-0.5$.  If we adopt the power of $-0.5$,
the final power of the delay time, $t$, in equation
(\ref{delay_time_simple_estimate_MS}) changes from $-1.0$ to $-1.2$

For the WD + RG system, we use $m= 5$ for the mass-luminosity relation
of the zero-age main-sequence stars with $M_2 = 0.7 - 2~M_\sun$
and $M_{2,0} \propto (t_2)^{-1/4}$.
Applying the area of the SN Ia regions that have been already
calculated as shown in Figure 12 of \citet{hkn99}, we numerically
obtain the same approximate relation as given by equation
(\ref{m10_power}) for $7 < M_{1,0} < 9$ and 
equation (\ref{m20_power}) for $0.9 < M_{2,0} < 2$.
Then we have the same power law index as in
equation (\ref{delay_time_simple_estimate_MS}), i.e.,
${\rm DTD}(t) \propto t^{-1}$ for the WD + RG channel.

In both the WD + MS and WD + RG channels, the DTD has
the power law index ($\propto t^{-n}$) close to $n=1$
regardless of the mass-lifetime dependence of 
$M_{2,0} \propto (t_2)^{-1/2.5}$ for the WD + MS 
or $M_{2,0} \propto (t_2)^{-1/4}$ for the WD + RG system.
The important ``details'' to realize such DTDs is how the SN Ia region
shrinks or expands as the initial secondary mass, $M_{2,0}$, decreases.
Here, the approximate relation given by equation(\ref{m20_power})
holds for both the WD + MS and WD + RG systems, which leads
to the logarithmic form of the DTD$(t) \propto \int d \log M_{2,0}
/ \Delta t ~\propto \int d \log t_2 / \Delta t ~\propto t^{-1}$
regardless of the mass-lifetime dependence, as seen in equation 
(\ref{delay_time_simple_estimate_MS}).

\section{Conclusions and Discussion}
\citet{tot08} argued that the ``detailed``
SD models on the basis of the population
synthesis should have prominent peaks in the DTD at characteristic time
scales of the secondary mass, thus being inconsistent with the
observation.  As already shown in Figure \ref{delay_time_4figures}a,
however, our DTD on the basis of the ``detailed''
binary evolution models has a featureless power law,
being in good agreement with the observation.
This is because the mass of the secondary star of the SN Ia
system ranges from $M_{2.0}
\sim 0.9$ to $6 ~M_\sun$ (Fig. \ref{zregevl10_strip_ms_rg}) due to
the effects of the WD winds and the mass stripping.
In our model, moreover, the number ratio of SNe Ia
between the WD + MS component and the WD + RG
component is $r_{\rm MS/RG} = 1.1$.  Such almost equal
contributions of the two components help to yield
a featureless power law as discussed below.

As for the metallicity effect \citep[e.g.,][]{kob08}, we assume
that the metallicity had already increased to $Z=0.02$ (or more)
at the birth of progenitor stars mainly because the galaxies
studied by \citet{tot08} consist of old galaxies,
the metallicity of which had already increased to $Z=0.02$ or more.
Therefore, Totani et al.'s (2008) data should not show any metallicity
effect even if the metallicity effect really exists.

In order to see the effect of different initial distributions
of the binary mass ratio $q$, we calculate the DTDs for two cases
of $f(q)$ \citep[e.g.,][]{gre83, gre05, kob08}.  
Figure \ref{delay_time_4figures}b shows that the DTD for
$f(q)= 2 / (1+q)^2$ is in good
agreement with the observation.  The other DTD for
$f(q)= 24 q^2/(1+q)^4$ shown in Figure \ref{delay_time_4figures}c
is marginally consistent with the observation.

Such a weak dependence stems from the fact that the ratio of
the two SN Ia components, $r_{\rm MS/RG}$,
depends slightly on $f(q)$.  The WD + MS component has a
relatively short delay time, and tends to have a large $M_{2,0}$.
In contrast, the WD + RG component has a long delay time,
having a small $M_{2,0}$.  In Figure \ref{zregevl10_strip_ms_rg},
for example, $M_{1,0} \sim 7 ~M_\sun$
for $M_{\rm WD,0} \sim 1 ~M_\sun$, while
$M_{2,0} \sim 0.9 - 3 ~M_\sun$ for the WD + RG system, i.e., $q \sim
0.13 - 0.4$.  The $q$ distribution of $f(q)= 24 q^2/(1+q)^4$ has a
peak at $q=1$ and takes smaller values at smaller $q$.
As a result, $r_{\rm MS/RG}= 1.4$ for this $f(q)=24 q^2/(1+q)^4$
is slightly larger than $r_{\rm MS/RG}= 1.0$ for $f(q)= 2 / (1+q)^2$.
Thus the DTDs of our ``detailed'' SD models
is not so sensitive to the mass ratio
distribution $f(q)$, suggesting that almost a featureless
power law shape ($\propto t^{-1}$, $n \approx 1$) from 
$t \sim 0.1$ to 10 Gyr is common among our ``detailed'' SD models
as long as the mass range of the secondary is 
$M_{2,0} \sim 0.9 - 6~M_\sun$.

We also calculate a DTD with a different power law index of
IMF, i.e., $\alpha = 2.35$
instead of $\alpha = 2.5$ in equations
(\ref{realization_frequency}) and (\ref{delay_time_distribution}).
As shown in Figure \ref{delay_time_4figures}d,
the results are hardly affected by changing $\alpha$ of the IMF
as long as $\alpha = 2.35 \pm 0.15$.

\acknowledgments
This research has been supported by World Premier International
Research Center Initiative (WPI Initiative), MEXT, Japan, and by
the Grant-in-Aid for Scientific Research of the Japan Society for the
Promotion of Science (18104003, 18540231, 20540226, 20540227) and MEXT
(19047004, 20040004).

\begin{deluxetable}{llll}
\footnotesize
\tablecaption{Delay time distribution of SNe Ia\tablenotemark{a}
\label{sn1a_delay_time_z02_const}}
\tablewidth{0pt}
\tablehead{
\colhead{delay time} & \colhead{WD + MS\tablenotemark{b}} 
& \colhead{WD + RG\tablenotemark{c}} & \colhead{total} \nl
\colhead{(Gyr)} & \colhead{} & \colhead{}
& \colhead{}
}
\startdata
0.05 & 0.0 & 0.0  & 0.0 \nl
0.1 & 3.85 & 0.0  & 3.85 \nl
0.2 & 1.56 & 0.0  & 1.56 \nl
0.4 & 0.617 & 0.0227 & 0.640 \nl
0.8 & 0.172 & 0.177  & 0.349 \nl
1.6 & 0.0  &  0.189  & 0.189 \nl
3.2 & 0.0  &  0.0734 & 0.0734 \nl
6.4 & 0.0  &  0.0286 & 0.0286 \nl
12.8 & 0.0 &  0.00913 & 0.00913 \nl
25.6 & 0.0 &  0.000417 & 0.000417
\enddata
\tablenotetext{a}{binary mass ratio distribution of $f(q)=1$,
IMF of $\alpha = 2.5$, and metallicity of $Z=0.02$}
\tablenotetext{b}{${\nu}_{\rm WD + MS} = 0.0035$~yr$^{-1}$ in our
Galaxy}
\tablenotetext{c}{${\nu}_{\rm WD + RG} = 0.0032$~yr$^{-1}$ in our
Galaxy}
\end{deluxetable}










\begin{thebibliography}{}
\bibitem[Belczynski et al. (2005)]{bel05}
Belczynski, K., Bulik, T., \& Ruiter, A. J. 2005, \apj, 629, 915

\bibitem[Belczy\'nski  \&  Miko{\l}ajewska  (1998)]{bel98}
Belczy\'nski, K., \& Miko{\l}ajewska, J. 1998, \mnras, 296, 77

\bibitem[Greggio (2005)]{gre05}
Greggio, L. 2005 \aap, 441, 1055

\bibitem[Greggio \& Renzini (1983)]{gre83}
Greggio, L., \& Renzini, A. 1983 \aap, 118, 217

\bibitem[Hachisu \& Kato (2003c)]{hac03kc}
Hachisu, I., \& Kato, M. 2003c, \apj, 598, 527

\bibitem[Hachisu et al. (2000a)]{hkkm00}
Hachisu, I., Kato, M., Kato, T., \& Matsumoto, K. 2000a,
\apjl, 528, L97 

\bibitem[Hachisu et al. (1996)Hachisu, Kato, \& Nomoto]{hkn96}
Hachisu, I., Kato, M., \& Nomoto, K. 1996, \apj, 470, L97 

\bibitem[Hachisu et al. (1999a)Hachisu, Kato, \& Nomoto]{hkn99}
Hachisu, I., Kato, M., \& Nomoto, K. 1999a, \apj, 522, 487 

\bibitem[Hachisu et al. (2008)Hachisu, Kato, \& Nomoto]{hkn08}
Hachisu, I., Kato, M., \& Nomoto, K. 2008, \apj, 679, 1390

\bibitem[Hachisu et al. (1999b)]{hknu99}
Hachisu, I., Kato, M., Nomoto, K., \& Umeda, H. 1999b, \apj, 519, 314


\bibitem[Hillebrandt \& Niemeyer (2000)]{hil00}
Hillebrandt, W., \& Niemeyer, J. 2000, \araa, 38, 191

\bibitem[Iben \& Tutukov (1984)]{ibe84}
Iben, I., Jr., \& Tutukov, A. V. 1984, \apjs, 54, 335

\bibitem[Kobayashi \& Nomoto (2008)]{kob08}
Kobayashi, C., \& Nomoto, K., 2008, \apj, submitted (arXiv:0801.0215)

\bibitem[Livio (2000)]{liv00}
Livio, M. 2000, Type Ia Supernovae: Theory and Cosmology,
(Cambridge: Cambridge Univ. Press), 33 (astro-ph/9903264)

\bibitem[Mannucci et al. (2005)]{man05}
Mannucci, F., Della Valle, M., Panagia, N., Cappellaro, E.,
Cresci, G., Maiolino, R., Petrosian, A., \& Turatto, M. 2005,
\aap, 433, 807

\bibitem[Mannucci et al. (2006)]{man06}
Mannucci, F., Della Valle, M., \& Panagia, N. 2006, \mnras, 370, 773

\bibitem[Matteucci et al. (2006)]{mat06}
Matteucci, F., Panagia, N., Pipino, A., Mannucci, F., Recchi, S.,
\& Della Valle, M. 2006, \mnras, 372, 265

\bibitem[Meng et al. (2008)Meng, Chen, \& Han]{men08}
Meng, X., Chen, X., \& Han, Z. 2008, \mnras, submitted (arXiv:0802.2471)

\bibitem[Nomoto et al. (1997)Nomoto, Iwamoto, \& Kishimoto]{nom97}
Nomoto, K., Iwamoto, K., \& Kishimoto, N. 1997, Science, 276, 1378

\bibitem[Nomoto et al. (2000)]{nom00}
Nomoto, K., Umeda, H., Kobayashi, C., Hachisu, I., Kato, M., \&
Tsujimoto, T. 2000, in AIP Conf. Proc. Vol. 522:
Cosmic Explosions: Tenth Astrophysics Conference,
ed. S. S. Holt \& W. W. Zhang (American Institute of Physics), 35
(astro-ph/0003134)

\bibitem[Ruiz-Lapuente \& Canal (1998)]{rui98}
Ruiz-Lapuente, P., \& Canal, R. 1998, \apj, 497, L57

\bibitem[Totani et al. (2008)]{tot08}
Totani, T., Morokuma, T., Oda, T., Doi, M., \& Yasuda, N. 2008, \pasj, 
submitted (arXiv:0804.0909)

\bibitem[Umeda et al. (1999)]{ume99}
Umeda, H., Nomoto, K., Yamaoka, H., \& Wanajo, S. 1999, \apj, 513, 861

\bibitem[Yungelson \& Livio (2000)]{yun00}
Yungelson, L. R., \& Livio, M. 2000, \apj, 528, 108
\end{thebibliography}
\end{document}